\pageno=1                                      
%
%
%
\font\ninerm=cmr9
\font\eightrm=cmr8
\font\sixrm=cmr6
\font\ninei=cmmi9
\font\eighti=cmmi8
\font\sixi=cmmi6
\skewchar\ninei='177 \skewchar\eighti='177 \skewchar\sixi='177
\font\ninesy=cmsy9
\font\eightsy=cmsy8
\font\sixsy=cmsy6
\skewchar\ninesy='60 \skewchar\eightsy='60 \skewchar\sixsy='60

\font\ninebf=cmbx9
\font\eightbf=cmbx8
\font\sixbf=cmbx6
\font\ninett=cmtt9
\font\eighttt=cmtt8
\hyphenchar\tentt=-1 
\hyphenchar\ninett=-1
\hyphenchar\eighttt=-1
\font\ninesl=cmsl9
\font\eightsl=cmsl8
\font\nineit=cmti9
\font\eightit=cmti8
\newskip\ttglue
\def\tenpoint{\def\rm{\fam0\tenrm}%
  \textfont0=\tenrm \scriptfont0=\sevenrm \scriptscriptfont0=\fiverm
  \textfont1=\teni \scriptfont1=\seveni \scriptscriptfont1=\fivei
  \textfont2=\tensy \scriptfont2=\sevensy \scriptscriptfont2=\fivesy
  \textfont3=\tenex \scriptfont3=\tenex \scriptscriptfont3=\tenex
  \def\it{\fam\itfam\tenit}%
  \textfont\itfam=\tenit
  \def\sl{\fam\slfam\tensl}%
  \textfont\slfam=\tensl
  \def\bf{\fam\bffam\tenbf}%
  \textfont\bffam=\tenbf \scriptfont\bffam=\sevenbf
   \scriptscriptfont\bffam=\fivebf
  \def\tt{\fam\ttfam\tentt}%
  \textfont\ttfam=\tentt
  \tt \ttglue=.5em plus.25em minus.15em
  \normalbaselineskip=12pt
  \let\sc=\eightrm
  \let\big=\tenbig
  \setbox\strutbox=\hbox{\vrule height8.5pt depth3.5pt width0pt}%
  \normalbaselines\rm}
\def\ninepoint{\def\rm{\fam0\ninerm}%
  \textfont0=\ninerm \scriptfont0=\sixrm \scriptscriptfont0=\fiverm
  \textfont1=\ninei \scriptfont1=\sixi \scriptscriptfont1=\fivei
  \textfont2=\ninesy \scriptfont2=\sixsy \scriptscriptfont2=\fivesy
  \textfont3=\tenex \scriptfont3=\tenex \scriptscriptfont3=\tenex
  \def\it{\fam\itfam\nineit}%
  \textfont\itfam=\nineit
  \def\sl{\fam\slfam\ninesl}%
  \textfont\slfam=\ninesl
  \def\bf{\fam\bffam\ninebf}%
  \textfont\bffam=\ninebf \scriptfont\bffam=\sixbf
   \scriptscriptfont\bffam=\fivebf
  \def\tt{\fam\ttfam\ninett}%
  \textfont\ttfam=\ninett
  \tt \ttglue=.5em plus.25em minus.15em
  \normalbaselineskip=10pt 
  \let\sc=\sevenrm
  \let\big=\ninebig
  \setbox\strutbox=\hbox{\vrule height8pt depth3pt width0pt}%
  \normalbaselines\rm}
\def\eightpoint{\def\rm{\fam0\eightrm}%
  \textfont0=\eightrm \scriptfont0=\sixrm \scriptscriptfont0=\fiverm
  \textfont1=\eighti \scriptfont1=\sixi \scriptscriptfont1=\fivei
  \textfont2=\eightsy \scriptfont2=\sixsy \scriptscriptfont2=\fivesy
  \textfont3=\tenex \scriptfont3=\tenex \scriptscriptfont3=\tenex
  \def\it{\fam\itfam\eightit}%
  \textfont\itfam=\eightit
  \def\sl{\fam\slfam\eightsl}%
  \textfont\slfam=\eightsl
  \def\bf{\fam\bffam\eightbf}%
  \textfont\bffam=\eightbf \scriptfont\bffam=\sixbf
   \scriptscriptfont\bffam=\fivebf
  \def\tt{\fam\ttfam\eighttt}%
  \textfont\ttfam=\eighttt
  \tt \ttglue=.5em plus.25em minus.15em
  \normalbaselineskip=9pt
  \let\sc=\sixrm
  \let\big=\eightbig
  \setbox\strutbox=\hbox{\vrule height7pt depth2pt width0pt}%
  \normalbaselines\rm}
%
\def\headtype{\ninepoint}                 
\def\abstracttype{\ninepoint}             
\def\captiontype{\ninepoint}              
\def\footnotetype{\ninepoint}             
\def\refit{\it}                           
\font\chaptitle=cmr10 at 11pt             
\rm                                       

%
%
\parindent=0.25in                         
\parskip=0pt                              
\baselineskip=12pt                        
\hsize=4.25truein                         
\vsize=7.445truein                        
\hoffset=1in                              
\voffset=-0.5in                           

\newskip\sectionskipamount                
\newskip\aftermainskipamount              
\newskip\subsecskipamount                 
\newskip\firstpageskipamount              
\newskip\capskipamount                    
\newskip\ackskipamount                    
\sectionskipamount=0.2in plus 0.09in
\aftermainskipamount=6pt plus 6pt         
\subsecskipamount=0.1in plus 0.04in
\firstpageskipamount=3pc
\capskipamount=0.1in
\ackskipamount=0.15in
\def\sectionskip{\vskip\sectionskipamount}
\def\aftermainskip{\vskip\aftermainskipamount}
\def\subsecskip{\vskip\subsecskipamount} 
\def\firstpageskip{\vskip\firstpageskipamount}
\def\capskip{\hskip\capskipamount}

%
%
\nopagenumbers                            
\newcount\firstpageno                     
\firstpageno=\pageno                      
\newcount\chapno                          

\def\rightheadline{\headtype\phantom{\folio}\hfil\runningtitletext\hfil\folio}
\def\leftheadline{\headtype\folio\hfil\runningauthortext\hfil\phantom{\folio}}
\headline={\ifnum\pageno=\firstpageno\hfil
           \else
              \ifdim\ht\topins=\vsize           
                 \ifdim\dp\topins=1sp \hfil     
                 \else
                     \ifodd\pageno\rightheadline\else\leftheadline\fi
                 \fi
              \else
                 \ifodd\pageno\rightheadline\else\leftheadline\fi
              \fi
           \fi}

\def\bottomnumber{\hss\tenrm[\folio]\hss}
\footline={\ifnum\pageno=\firstpageno\bottomnumber\else\hfil\fi}

%
%
%
%
\outer\def\mainsection#1
    {\vskip 0pt plus\smallskipamount\sectionskip
     \message{#1}\vbox{\noindent{\bf#1}}\nobreak\aftermainskip\noindent}
 
\outer\def\subsection#1
    {\vskip 0pt plus\smallskipamount\subsecskip
     \message{#1}\vbox{\noindent{\bf#1}}\nobreak\smallskip\nobreak\noindent}
 
\def\backup{\nobreak\vskip-\baselineskip\nobreak\vskip-\subsecskipamount\nobreak
}

\def\title#1{{\chaptitle\leftline{#1}}}
\def\name#1{\leftline{#1}}
\def\affiliation#1{\leftline{\it #1}}
\def\abstract#1{{\abstracttype \noindent #1 \smallskip\vskip .1in}}
\def\ref{\noindent \parshape2 0truein 4.25truein 0.25truein 4truein}
\def\caption{\noindent \captiontype
             \parshape=2 0truein 4.25truein .125truein 4.125truein}

\def\footnote#1{\edef\fspafac{\spacefactor\the\spacefactor}#1\fspafac
      \insert\footins\bgroup\footnotetype
      \interlinepenalty100 \let\par=\endgraf
        \leftskip=0pt \rightskip=0pt
        \splittopskip=10pt plus 1pt minus 1pt \floatingpenalty=20000
        \textindent{#1}\bgroup\strut\aftergroup\strut\egroup\let\next}
\skip\footins=12pt plus 2pt minus 4pt 
\dimen\footins=30pc 

%
%

\def\@{\spacefactor 1000}

\def\,{\pcomma} 
\def\pcomma{\relax\ifmmode\mskip\thinmuskip\else\thinspace\fi}

\def\oversim#1#2{\lower0.5ex\vbox{\baselineskip=0pt\lineskip=0.2ex
     \ialign{$\mathsurround=0pt #1\hfil##\hfil$\crcr#2\crcr\sim\crcr}}}


\input epsf

\def\t0{\theta_{\circ}}

\def\be{\begin{equation}}
\def\en{\end{equation}}

\def\msun{M_{\odot}}
\def\rsun{R_{\odot}}
\def\lsun{L_{\odot}}
\def\msunyr{\msun yr^{-1}}

\def\mdot{\dot{M}}

\def\runningtitletext{EVOLUTION OF DISK ACCRETION}
\def\runningauthortext{CALVET ET AL.}

\null
\firstpageskip

{\baselineskip=14pt
\title{EVOLUTION OF DISK ACCRETION}
}

\vskip .3truein
\name{NURIA CALVET\footnote{*}{Also Centro de Investigaciones
de Astronom{\'\i}a} and LEE HARTMANN }
\affiliation{Harvard-Smithsonian Center for Astrophysics}
\vskip .2truein
\leftline{and}
\vskip .1truein
\name{STEPHEN E. STROM}
\affiliation{Five College Astronomy Department, University
of Massachusetts\footnote{**}{Now at
National Optical Astronomy Observatories}}
\vskip .3truein

\abstract{
We review the present knowledge of disk accretion
in young low mass stars, and in particular, the mass 
accretion rate $\mdot$ and its
evolution with time.  The methods used to obtain mass
accretion rates from ultraviolet excesses and emission
lines are described, and the current best estimates of $\mdot$ for Classical
T Tauri stars and for objects still surrounded by 
infalling envelopes are given. We argue that the low mass accretion rates 
of the latter objects require episodes of high mass accretion
rate to build the bulk of the star. 
Similarity solutions for viscous disk evolution suggest that
the inner disk mass accretion rates can be self-consistently
understood in terms of the disk mass and size if the viscosity
parameter $\alpha \sim 10^{-2}$.
Close companion stars may accelerate the disk accretion process, resulting
in accretion onto the central star in $\le$ 1Myr; this may help
explain the number of very young stars which are not currently
surrounded by accretion disks (the weak emission T Tauri stars). 
}


\mainsection{I.~~INTRODUCTION}
\backup

The initial angular momenta of star-forming molecular cloud cores must be
responsible for the ultimate production of binary (and multiple)
stellar systems and circumstellar
disks.  Unless the protostellar cloud core is very slowly rotating,
much or most of the stellar mass is likely to land initially on a disk,
and must subsequently be accreted from the disk onto the protostellar core.
In the early phases of this accretion, the circumstellar disks may be
relatively massive in comparison with the central protostar, and so
gravitational instabilities may be important in angular momentum transport 
and consequent disk evolution.  At later phases, disk evolution is likely
to be driven by viscous processes, perhaps limited by condensation of
bodies.  Stellar magnetic fields may ultimately halt the accretion
of material onto the central star.
There are substantial uncertainties in our theoretical understanding
of these processes, and we must rely on observations for guidance.

In this chapter we review the present knowledge of the disk accretion process
around low mass stars, 
in particular the rate at which mass is transferred onto the star, 
$\mdot$, and its evolution with time.  
Knowledge of $\mdot$ will help put lower limits on the disk
mass at a given age, independent of uncertainties in dust opacities.
It puts constraints on disk physics, and in particular on
temperatures and surface densities at a given age, and thus
on conditions which obtain during the time when solid bodies agglomerate.
We begin with
a description of the
different methods of determining disk accretion rates and
the values of $\mdot$ obtained applying these methods
to stars in different groups and environments.
We then discuss
the implications of the observational results for disk
accretion physics and evolution, and in particular investigate whether
the data are consistent with simple models of viscous disk evolution.

\mainsection{II. DETERMINATION OF $\mdot$}
\backup
\subsection{A.~~Infrared excesses}

When the infrared excess emission in 
Classical T Tauri stars (hereafter CTTS) was recognized as emission from
dust at low temperatures distributed in a circumstellar disk,
it was thought that the excess energy would be a direct measurement
of the accretion luminosity ($L_{acc} = G {\mdot} M_*/R_*$,
where $M_*$ and $R_*$ are the stellar mass and radius). However, it soon became
apparent that their spectral energy distributions (SEDs) did not
follow the law $\lambda F_{\lambda} \propto {\lambda}^{-4/3}$ expected for
standard accretion disks, but were much flatter (Rydgren and Zak 1987; Kenyon and Hartmann 1987).
It has now been realized that the major agent heating the disks of many
CTTS is irradiation by the central star (Adams and Shu 1986;
Kenyon and Hartmann 1987; Calvet et al. 1991, 1992; Chiang and Goldreich 1997; 
D'Alessio et al. 1998).
The optically-thick disks of CTTS are probably ``flared'', i.e., have ``photospheres''
that curve away from the disk midplane; this makes them more efficient in
absorbing light from the central star; this extra heating helps to increase  
the disk scale height and thus increases the flaring. Self-consistent
calculations (in various approximations) indicate that the flaring is especially
important at large radii
(Kenyon and Hartmann 1987; Chiang and Goldreich 1997; D'Alessio et al. 1998).
This makes it extremely difficult, if not impossible, to say anything about accretion
energy release, and thus accretion rates, in outer disk emission.

For many CTTS it is very difficult
to extract quantitative estimates of
mass accretion rates even from emission from the inner, flat parts of the disk.
The basic reason is that the accretion luminosities of
many CTTS are smaller than the luminosity the optically-thick inner
disk produces as a result of absorbing light from
the central star.

The effective temperature $T$ of the 
disk is determined by internal viscous
dissipation and external irradiation by the central star;
at a given annulus, 
$$T^4(r) \sim {T_v}^4 + {T_i}^4 , \eqno(1) $$
\noindent
where $T_v = { [ ( 3G M_* \mdot / 8 \pi \sigma r^3) f ] }^{1/4}$ is the effective
temperature that would result from accretion without irradiation,
with $ f =  1 - \sqrt {(R_*/r)}$,
and ${T_i}^4 = (2 / 3 \pi ) {T_*}^4 (R_*/r)^3  $ is the 
effective temperature for the case of only irradiation
(the flat disk approximation). 
In these expressions, 
we assume that the mass accretion rate through the inner disk
is constant and equal to the rate  at which mass is transferred onto the star.
The mass accretion rate
at which $T_i \sim T_v$ is 
$$ {\mdot}_{c} \sim 2 \times 10^{-8} \msunyr
(T_*/4000 K)^4 (R_*/2 \rsun)^3 
(M_*/0.5 \msun)^{-1},$$
\noindent calculated for
typical Taurus CTTS parameters (K7-M0, age $\sim$ 1 Myr) and $ r \sim 3 R_*$. 
Since the median value of $\mdot$ in CTTS is 
$\sim 10^{-8} \msunyr$ (see below), 
this implies that irradiation dominates the inner disk heating
for a large number of the stars
and determines the amount of flux excess. Only for objects with 
significantly high $\mdot$ will the (near-infrared) disk emission
depend upon the mass accretion rate.

Other considerations affect a disk's near-infrared emission.
It was originally thought that CTTS disks extended all the way into the
star, and disk material joined the star through a ``hot boundary layer'',
where half of the accretion energy was dissipated, while the other
half was emitted in the disk (cf. Bertout 1989). It has now become
apparent that for typical values of the stellar magnetic field
($< 1$ kG, Basri et al. 1992) and
of the mass accretion rate, 
the inner disk of CTTS can be disrupted by the magnetic
field (K\"onigl 1991; Najita et al., this volume); material falls onto 
the star along the magnetic field
lines, forming a magnetosphere, and merges with the star
through an accretion shock at the stellar surface.
In support of this model,
fluxes and line profiles of the broad (permitted) emission lines in CTTS
are well explained by the magnetospheric flow (Calvet and Hartmann 1992;
Hartmann et al. 1994; Edwards et al. 1994;
Muzerolle et al. 1998a,b,c; Najita et al., this volume), while the ultraviolet
and optical excess fluxes are well accounted for by the accretion shock emission
(Calvet and Gullbring 1998).  

For truncation radii
$R_i \sim 3 - 5 \, R_*$, the maximum
temperature in a disk with typical parameters will be $\sim 1200 - 850 K$,
so that disk emission drops sharply below $\sim 2 - 3.5 \mu$m.
The calculation of the disk luminosity from the observed excess also depends
upon the cosine of the inclination angle, which is generally unknown.
As an illustration of these points,
Meyer et al. (1997) have shown that the so-called
``CTTS locus'' in JHK and HKL diagrams, i.e., the region 
populated by CTTS outside
that corresponding to reddened main sequence stars
(Meyer et al., this volume)
could be explained
in terms of emission by irradiated accretion disks with inner holes
of different sizes (but inside co-rotation radii), and a random
distribution of inclinations. However, no one-to-one correlation
could be found between the excess and $\mdot$.
In a sample of stars in Taurus with known reddening and $\mdot$,
the largest excesses in the
locus were produced by stars with the highest $\mdot$,
but the converse was not true, due to the
effects of holes and inclination.   

Finally, in objects surrounded by 
infalling dusty envelopes, emission from the 
dust destruction radius, which peaks at $\sim 2 \mu$m,
may contribute significantly to the K band and longward 
(Calvet et al. 1997), hiding the intrinsic disk emission.

The difficulty of deriving reliable measures of $\mdot$ from
excess infrared emission requires using alternative methods
of estimating $\mdot$'s. We will consider such methods in 
the next two sections.

\subsection{B.~~Veiling luminosities}

The best evidence for CTTS disk accretion generally comes from
the interpretation of the ultraviolet and optical spectra.
The photospheric absorption lines are ``veiled'',
that is, they are less deep than standard stars of the same spectral type.
This veiling is produced by (mostly) continuum
emission from a region hotter than the stellar photosphere.
The luminosity of the veiling or excess continuum in CTTS
is typically $\sim$ 10 \% of the total stellar luminosity. It is difficult
to account for this extra energy with a stellar origin, and
impossible in the case of most extreme
CTTS, where the excess continuum luminosity is several times the stellar
luminosity.  This conclusion is reinforced by the existence of the
weak-emission T Tauri stars or WTTS.  The WTTS have similar masses and
ages as the CTTS in many regions, but do not exhibit the UV-optical
veiling continuum of CTTS, showing that the excess emission is not
an intrinsic property of young stars.  The veiling
emission is observed only when there is excess near-infrared emission
(Hartigan et al. 1995; Najita et al., this volume),
strongly supporting the idea
that accretion from a disk is occurring, producing both the infrared
excess as well as the hot continuum, as originally envisaged by
Lynden-Bell and Pringle (1974). 

In the magnetospheric model, the disk is truncated in the inner regions,
which is precisely where an accretion disk would emit most of its 
energy.  The total disk accretion luminosity is expected to be about
$$L_{acc} (disk) \sim G M_* \mdot /(2 R_i) +  L_{diss}\,$$
where $L_{diss}$ is any energy that might be dissipated in
the disk by the stellar magnetic field lines (Kenyon et al. 1996).
If we assume that the effect of the stellar magnetic field is to
substantially reduce the angular momentum of the disk material, so
that it starts out nearly at rest at $R_i$ before falling onto the
star along the magnetic field lines, then the infalling material
should dissipate its energy at the stellar surface at a rate
$\sim (G M_* \mdot /R_* )(1 - R_*/R_{i})$.  
For disk truncation radii $\sim 3 - 5 R_*$,
most of the accretion energy is released at the stellar
surface in the hot accretion shock whose radiation
is observed as the veiling continuum (K\"onigl 1991; Calvet and
Gullbring 1998). 

Accretion luminosities can be obtained from
measurements of the veiling of the absorption lines
by the following procedure in which excess and
intrinsic photospheric emission components are separated.
The observed fluxes are fitted by the scaled, dereddened fluxes of a
standard star, assumed to be of the same spectral type 
as the object star,
plus a continuum flux, which produces the observed veiling 
at each absorption line wavelength. This fit
yields the spectrum of the excess continuum and the reddening towards
the star.  The total
excess luminosity is calculated from the measured luminosity 
using a model to extrapolate the emission to unobserved wavelengths.
Finally, the mass accretion rate can be calculated from stellar
mass and radius estimated from the location of the star 
in the HR diagram and
comparison with evolutionary tracks.

Accretion luminosities for CTTS in Taurus
have been determined from veiling measurements by
a number of authors, including Hartigan et al. (1991), 
Valenti et al.
(1993), and Hartigan et al. (1995). 
The mass accretion rates derived from these studies 
range from
$\sim 10^{-7} \msunyr$ 
for Hartigan et al., to
$\sim 10^{-8} \msunyr$ 
for Valenti et al.
More recently, Gullbring et al. (1998, GHBC), re-measured accretion
rates for a sample of Taurus CTTS using
spectrophotometry covering
the blue and Balmer jump region of the spectrum
down to the atmospheric cutoff, and found their results to
agree with the previous lower estimates.
In a sample of 17 CTTS in Taurus, 
GHBC found a median value of $\mdot$
$\sim 10^{-8} \msunyr$,
with a factor of $\sim$ 3 in estimated error.

These differences reflect the cumulative effects of
a number of differences in working assumptions
rather than a fundamental difference in approach.
Among them we can list the following:
(1) Adopted evolutionary tracks;
(2) Adopted model of accretion. The magnetosphere 
model predicts more luminosity per mass
accretion rate than the old boundary layer model; 
(3) Adopted physics. 
Some treatments assumed that a substantial fraction of the
emitted accretion energy was absorbed by the star,
which seems unlikely (Hartmann et al. 1997);
(4) Differing methods used to determine extinction corrections.
The photospheres of T Tauri stars show
color anomalies relative to standards, which render the determination
of reddening uncertain. In particular, GHBC found large color anomalies
in some non-accreting WTTS often used as templates; this would cause
the extinction to be overestimated in all objects.
These anomalies are probably due to spots on
the stellar surface and/or unresolved, cooler companion stars. 

To provide a powerful tool to determine mass accretion rates
for large samples of stars for which short wavelength 
spectrophotometry is difficult to obtain,
GHBC determined the relationship between the accretion 
luminosity and the excess luminosity in the U photometric band, $L_U$,
for the stars in their sample.
Figure 1a shows the excellent relationship between log $L_{acc}$ and log $L_U$. 
A least-squares fit to the line yields:
$$log \, (L_{acc}/L_{\odot}) = {1.04}^{+0.04}_{-0.18} \, log \, (L_{U}/L_{\odot})
        + {0.98}^{+0.02}_{-0.07} \eqno(2)$$
The spectrophotometric sample from which equation (2) was derived
is formed by stars in the K5-M2 range, with most of the stars
being K7-M0. The application of this calibration to
a wider spectral type/mass range remains to be confirmed, but
theoretical models of accretion shock emission from stars
of differing mass and radii indicate that for
the characteristic energy fluxes found in the accretion columns
of CTTS, the spectrum of the excess emission does not depend
on the underlying star, and the proportion of the total excess luminosity
in the U band ($\sim$ 10\%) displayed by equation (2) holds
(Calvet and Gullbring 1998).

Mass accretion rates have been determined for a larger sample of TTS
in Taurus using equation (2) by Hartmann et al. (1998, HCGD).
The median mass accretion rate is $\sim 10^{-8} \msunyr$, 
similar to that of the spectrophotometric sample, but
the errors are larger, given the lack of simultaneity in the photometric
measurements and the high degree of variability of CTTS.
HCGD also used spectral types
and photometry from the compilation of Gauvin and Strom (1992) to
estimate mass accretion rates for K5-M3 stars in the Chamaeleon I association. 
The median mass accretion rate
in Chamaeleon I is $\sim 4 \times 10^{-9} \msunyr$.
A histogram of the mass accretion rate for TTS in
Taurus and Chamaeleon I determined from the ultraviolet
excess is presented in Fig.\ 2a.

\subsection{C.~~Magnetospheric emission lines}
The measurement of mass accretion rates from the optical and ultraviolet
excess fluxes, either spectrophotometrically or from broad-band photometry, 
is sensitive to reddening corrections.  In heavily-extincted stars,
the UV fluxes are either very uncertain or unobservable; thus,
in early stage of evolution such as the infall/protostellar phase,
or in very young, dense regions of star formation,
such methods cannot be used.  For these objects, it is necessary
to devise methods to measure $\mdot$ 
in spectral regions that are less affected by intervening dust.

Most emission lines present in the spectra of young objects are
thought to be produced in the magnetospheric flow
(Najita et al., this volume).
Since the material flows through the magnetosphere with a
rate similar to $\mdot$ in the inner disk,
the emission fluxes of the lines formed in this flow are expected to 
depend upon the mass accretion rate.
Theoretical models show this to be the case, but other
parameters, such as the unknown temperature structure and the
characteristic size play a role too, as well as the
optical depth of the line (Muzerolle et al. 1998a).
For these reasons, empirical correlations between line luminosities and
accretion luminosities have been investigated, leaving the interpretation
to future theoretical work.

Muzerolle et al. (1998b,c, MHCb,c) 
have undertaken spectroscopic
studies in the red and infrared of the sample of stars
with accretion rates determined from spectrophotometric
measurements, and have found remarkably good correlations
between the luminosity of the Ca II triplet 8542, Pa$\beta$,
and Br$\gamma$ lines with accretion luminosities.
Figure 1b show the correlation between $L_{acc}$ and
the luminosity in Br$\gamma$, as an example.  Least-squares fits
to the data give:
$$       log \, (L_{acc}/L_{\odot}) = (0.85 \pm 0.12) \, log \, (L_{Ca II 8542}/L_{\odot}) 
        + (2.46 \pm 0.46) \;, \eqno(3)$$
$$       log \, (L_{acc}/L_{\odot}) = (1.03 \pm 0.16) \, log \, (L_{Pa\beta}/L_{\odot})
        + (2.80 \pm 0.58) \;, \eqno(4)$$
\noindent
and
$$       log \, (L_{acc}/L_{\odot}) = (1.20 \pm 0.21) \, log \, (L_{Br\gamma}/L_{\odot})
        + (4.16 \pm 0.86) \;.  \eqno(5)$$
These subsidiary calibrators of the accretion rate provide the means
to determine accretion rates in heavily extincted objects for the first time.

MHCc used luminosities of the Br$\gamma$ line for extincted Class II sources
in $\rho$ Oph from Greene and Lada (1996) to make the first determination
of accretion luminosities for these sources. They
find that the distribution of accretion luminosities is similar to
that in Taurus. Using approximate spectral types from Greene and Meyer (1995),
the estimated median mass accretion rate is $\sim 1.5 \times 10^{-8} \msunyr$,
although the span of spectral types covered is wider than in the
Taurus sample.

Even more far reaching, 
MHCc determined accretion luminosities for the deeply embedded Class I objects
(sources still surrounded by infalling envelopes) for the first time,
using Br$\gamma$ luminosities. 
Since typically $A_V \sim 20 - 30$ for Class I objects,
significant extinction is expected at Br$\gamma$ ($A_K \sim 2 - 3$),
which introduces large uncertainties in the determination of
the line luminosities. The light from the central star+disk
and from the envelope itself
is absorbed {\it and} scattered by the infalling envelope, in
proportions that depend on uncertain parameters such as the
geometry of the inner envelope (Calvet et al. 1997). 
Following Kenyon et al. (1993b),
MHCc calculate a correction factor K-K0, by assuming that the
central object has J-K colors similar to CTTS, and estimating the
intrinsic J magnitude from the bolometric luminosity.

Figure 3 shows accretion luminosities estimated by MHCc for a small number
of Class I sources in Taurus
and in $\rho$ Oph which have data on Br$\gamma$ 
(Greene and Lada 1996) plotted against the
total luminosities. The accretion luminosities of Class I
sources are significantly lower than the system
bolometric luminosity; in fact, the mean accretion
luminosity is  $\sim$ 10 - 20\% of the
mean bolometric luminosity of the sample.
This fact implies that the luminosity of Class I is dominated by the stellar component,
and gives a natural explanation to the similarity of the distributions of luminosities
of Class I and optically visible T Tauri stars in Taurus
(Kenyon et al. 1990).

Determining the mass accretion rate from $L_{acc}$ requires knowledge of
$M_*/R_*$.
We can obtain this ratio assuming that the stars are still on the birthline
(Stahler 1988), 
which seems justified since they are still accumulating mass from the envelope/disk.
We used the observed bolometric luminosity to locate the object
along the birthline (calculated for an infall rate of 
$2 \times 10^{-6} \msunyr$).
Figure 2b shows the distribution of ${\mdot}$ thus obtained 
for Class I sources in Taurus and $\rho$ Oph.

The mean mass infall rate in the envelope, ${\mdot}_i$, has been
estimated in Taurus from fitting the spectral energy distributions and
scattered light images of Class I sources
to be $\sim 4 \times 10^{-6} \msunyr$ (Kenyon et al. 1993a),
a value which is consistent with theoretical expectations (Shu et al. 1987).
The present determination of disk accretion rate 
shown in Fig.\ 2b indicates
that many Class I objects
are slowly accreting from their disks, despite the fact that mass is being deposited
in the outer disk at a higher rate.
This discrepancy was first recognized by Kenyon et al. (1990),
as the so-called luminosity problem of Class I sources.
If infall were spherical, the luminosity of the system
would be given by an accretion
luminosity $\sim G M_* {\mdot}_{i} /R_*$.
But in this case, 
the predicted accretion luminosity should be
of order $10 \lsun$, while the mean luminosity of Taurus
Class I sources is $\sim 1 - 2 \lsun$ (Kenyon et al. 1990, 1994).
Kenyon et al. (1990) pointed out that the mass accretion
rate in the disk onto the star does not necessarily equal the mass infall
rate from the envelope onto the disk, since they are
regulated by different physical processes.
The imbalance between the infall and accretion rates in this picture
leads to an accumulation of mass in the disk.  These disks could
eventually become gravitationally unstable (Larson 1984), with consequent rapid
accretion until sufficient mass has been emptied out of the disk.
Kenyon et al. (1990) suggested that these episodes could be related to 
the FU Orionis outbursts.

\subsection{D.~~The FU Ori disks}

The FU Orionis outbursts are now recognized as a transient
phase of high mass accretion in the disk around a forming
low mass star (see review by Hartmann and Kenyon 1996 and
references therein).
Although so far FU Ori objects have been mostly studied as isolated
phenomena, it is becoming increasingly clear
that episodes of high mass accretion rate may be a crucial,
if not dominant, process in the formation of stars.
We briefly review here the determination of $\mdot$
for these objects.

The canonical FU Ori objects were discovered from their increase
in brightness by several magnitudes over time scales of months
to years (Herbig 1977), during which the
luminosity increased from values typical to CTTS to a few hundred
$\lsun$. Since the emission
is dominated by the accretion disk in FU Ori objects, 
the accretion luminosity can be readily determined
from the observed SED.
Additional information is required to independently obtain $M_*/R_*$
and $\mdot$,
which can be estimated from
the surface temperature in the inner disk ($R \sim $ 2 - 3 $R_*$),
$T_{max} \sim 7000 K {(L_{acc}/100 \lsun )}^{1/4}
{(R_*/2 \rsun)}^{1/2}$ (from $T_v$ in equation [1]),
and measurements of the rotational velocity.
Using this method, $\mdot \sim 10^{-4} \msunyr $
has been inferred for the canonical objects for which
outbursts have been observed, consistent with their
high luminosities, $> 100 \lsun$.

A significant number of objects have been identified as 
FU Ori objects in recent
years, mostly at IR wavelengths. Photometric variability indicative of outbursts has been
detected in only a few objects;  their classification as FU Ori systems
has been based mostly on the presence of the near-infrared first overtone bands
of CO in deep absorption, comparable to late type giants
and supergiants and to the canonical FU Ori objects. Most of these objects are
in very early stages of evolution, as is their prototype L1551 IRS 5,
embedded in infalling envelopes
and associated with Herbig-Haro objects, jets and/or molecular outflows
(Elias 1978, Graham and Frogel 1985, Reipurth 1985; Staude and Neckel 1991;
Kenyon et al. 1993; Hanson and Conti 1995; Hodapp et al. 1996;
Reipurth and Aspin 1997; Sandell and Aspin 1998).
The luminosities of the objects identified
so far as members
of the FU Ori class range from $\sim 10 \lsun$ to $800 \lsun$. 
Since $\mdot \sim 2 \times 10^{-6} \msunyr (L_{acc}/10 \lsun )
$$\times { [ (M_*/R_*) / 0.18 \msun/\rsun ]} ^{-1}$, 
this range 
implies $\mdot \ge $ few $ \times 10^{-6} \msunyr$
(assuming they are on the birthline,
with typical $M_*/R_* \sim 0.18 \msun/\rsun$).
This value is consistent with the presence of CO in absorption,
since disk atmosphere calculations indicate that
the temperature of the continuum forming region 
is higher than the surface temperature (neglecting any 
wind contribution) for $\mdot > 10^{-6} \msunyr $ 
(Calvet et al. 1992).

The nature of the low luminosity FU Ori objects
and their relationship to the canonical objects remains to 
be elucidated. One possibility
is that disks undergo instabilities driving
outbursts of different magnitude.
Alternatively, these objects
could be in the phase of decay from a canonical FU Ori outburst.
An argument in favor of the second possibility comes 
from comparison of the mass loss rate required to drive the
molecular outflow and the present mass accretion rate. For L1551 IRS5
and PP13S ($L = 10 \lsun$ and $30 \lsun$, respectively),
the momentum flux of the molecular flow is $\sim$ few $\times 10^{-4} \msunyr {\rm km s^{-1}}$.
(Moriarty-Schieven and Snell 1988; Sandell and Aspin 1997).
For typical velocities of the jet of few $ \times 100 \, {\rm km s^{-1}}$, and
assuming momentum conservation, the 
momentum flux implies mass loss rates of the same order to the
inferred mass accretion rates, while both theoretically and observationally,
the ratio between the mass loss rate and mass accretion rate is close to $\sim 0.1$
(Calvet 1997).
This discrepancy may imply that the mass accretion rate of the disk
was much higher when the material giving rise to the molecular
outflow was ejected.

\mainsection{III. ~~THE EVOLUTION OF ACCRETION}
\backup

\subsection{A.~~ Observed $\mdot$ vs. age}

Figure 4 shows mass accretion rate vs. age for CTTS in
Taurus, $\rho$ Oph, and Chamaeleon I, for which
significant infall from an envelope has stopped.
We also show the 
range of $\mdot$ covered by 
Class I sources in Taurus, assuming a median age of 0.1 Myr
(estimated from the ratio of the number of Class I sources 
to T Tauri stars in Taurus, and adopting a mean age of 1 Myr 
for the latter, Kenyon et al. 1990). 
The data indicates a clear trend of accretion decaying with time
(HCGD; see also Hartigan et al. 1995),
even in the relatively short age spread covered by the observations.
There is also a very large spread in mass accretion rate at a given age,
which makes it very difficult to quantify the rate of decay.
HCGD shows that the slope of a least-squares-fit to the data
is highly dependent on the errors of $\mdot$ and age. If errors
are assumed larger in $\mdot$ than in age (the most likely situation, cf. HCGD),
then the slope is $\approx -1.5$ (with large uncertainty).

\subsection{B.~~ Disk Masses}

We can use the observed mean decay of $\mdot$ with time to 
obtain an estimate for the mass of the disk
for comparison with masses estimated from dust emission.
The amount of mass accreted
by a CTTS from the present time to infinity constitutes a lower limit to the 
mass remaining in 
the disk. If $\mdot(t) ~\sim \mdot(t_o) (t/t_o)^{-\eta}$, then this
limit to the disk mass is $M_{acc} ~=~ {\mdot(t_o) t_o / (\eta ~-~1)}
\sim 2 \mdot(t_o) t_o$, with $\eta \sim 1.5$. This gas mass can be compared to the disk
mass estimated from dust emission in the submillimeter and millimeter range,
$M_{mm}$, which is very dependent on the assumed opacities 
(Beckwith et al. 1990; Osterloh and Beckwith 1995, masses corrected by a 
factor of 2.5, HCGD).
The comparison (for single stars) yields $log (2 \mdot \times age/M_{mm}) 
= -0.07 \pm 0.21$, 
indicating that masses inferred from the current mass accretion rates and 
age estimates, which are probing the gas, are 
consistent with disk masses estimated independently from mm-wave dust opacities,
which in turn suggests that the dust opacities used in the latter estimates
are appropriate.

\subsection{C.~~ Early stages}

Figure 4 indicates that the mean mass accreted onto the star 
during the CTTS phase is $\sim 10^{-2} \msun$. 
This suggests that by the time stars reach the optically
visible CCTS stage, the remaining disk mass is relatively low
and little mass is added to the forming star.
This picture implies that the bulk of stellar
accretion must occur during the (highly-extincted)
infall phase, when the disk is being continually replenished
by the collapsing envelope.  
However, during the infall phase, our results suggest
that disk accretion rates during quiescent phases are 
only slightly higher than
those in the CTTS stage, 
precluding addition of more than $\sim 0.01 \msun$ during
these phases.
Significant additions to the mass of the growing
protostar must therefore come via material
accreted during transient episodes of high accretion.

Andr\'e et al. (1993) suggest that the Class 0 objects
are in the earliest phases of envelope collapse, when central accretion
rates are expected to be the highest, and argue 
that the Class 0 sources are
the true protostars.  
These very heavily extincted objects
have somewhat higher luminosities than the mean Class I luminosity
in Taurus,
and so may have higher accretion rates onto the central star.
However, Class 0 sources tend to be less frequent than Class I sources,
especially in Taurus; thus, the lifetime of the Class 0 phase may be
too short to account for most of the stellar accretion.  

The high accretion rate episodes required to explain the accretion
of the stellar mass can be attributed to 
instabilities in quiescent low $\mdot$ disks,
which result in outbursts and transient periods of high mass accretion rates.  
Several models have been presented to trigger those outbursts.
The most accepted model
is that of thermal instabilities in the inner disk (Lin and Papaloizou 1985;
Clarke et al. 1990; Kawazoe and Mineshige
1993; Bell and Lin 1994; Bell et al. 1995).
This model can naturally explain the occurrence of outbursts
during the infall phase because it requires a 
high background mass accretion rate
from the outer disk, of the order of a few $\times 10^{-6} \msunyr$,
to match the observations.
Moreover, according to this model outbursts will be triggered in the disk 
as long as mass is deposited in the
outer disk at this rate, ensuring that mass will be transferred
from the initial cloud into the star.

The number of currently-known FU Ori objects is insufficient to explain 
the formation of typical stars mostly through outbursts (Hartmann and Kenyon 1996).
However, the known sample of FU Ori disks is very incomplete, because
the identifying characteristic of high mass accretion rate
disks is the near-infrared bands of CO in absorption and
many objects may be too heavily extincted to be detected at $\sim 2.2 \mu$m.
More and more sensitive observations of embedded
objects in the near-infrared are necessary to test this hypothesis.

\subsection{D.~~ Viscous evolution in TTS disks}

Once the main infall phase is over and the envelope is no
longer feeding mass and angular momentum into the disk, 
we expect disk evolution to be driven mostly by viscous processes,
namely, those in which the angular momentum transport is provided 
by a turbulent viscosity. In this and the next section, we attempt
to interpret the observed properties of CTTS disks in terms 
of viscous evolution, with the aim of understanding the
main physical processes at play  and
the role of initial and boundary conditions. Similarly,
physical models for disk evolution help us relate 
properties and evolution of the inner disk, as measured
by ${\mdot} (t)$, to those of the
outer disk, such as radius and mass, $R_d$ and $M_d$.

The disk angular momentum is 
$$J_d ~=~ \int_0^{M_d} dM \, \Omega R^2 ~  ~\propto~ M_d R_{d}^{1/2}.\eqno(6)$$
\noindent
where $\Omega$ is the Keplerian angular velocity.
If we neglect the (small amount) of angular momentum being added to the star,
or the possible angular momentum loss to an inner disk wind (Shu et al. 1994;
Shu et al., this volume), then the disk angular momentum is approximately constant.
In turn, this requires  $R_d \propto {M_d}^{-2}$, 
so angular momentum conservation implies that the
disk expands as the mass of the disk is accreted to the star.
Evolution occurs on the viscous time scale,
$t_{visc} \sim R^2 / \nu$, where $\nu$ is the viscosity (Pringle 1981).
If $\nu \propto R^{\gamma}$, then $dR_d / dt \sim R_d/ t_{visc} \propto {R_d}^{\gamma -1}$,
so 
$R_d \propto t^{ 1 /( 2 - \gamma) }$,
$M_d \propto t^{-{ 1/ 2 ( 2 - \gamma) }}$, and
$\mdot \propto t^{-{ ( 5/2 - \gamma ) / ( 2 - \gamma) }}$.
Therefore and in principle, from the observed decay of $\mdot$ with time, 
$\mdot(t) ~\propto  t^{-\eta}$, we can obtain
$\gamma = (2 \eta - 5/2) / (\eta - 1)$, and
the evolution of radius and mass of the disk can be predicted.

As Lynden-Bell and Pringle (1974) showed, analytic similarity solutions
describing the evolution of disk properties
exist for the case of power-law viscosity (see also Lin and Bodenheimer 1982).
These analytical similarity solutions have been applied by HCGD
as a first approximation to the evolution of T Tauri disks. 
(More complex models have been used by Stepinski (1998) to consider
similar issues; whether the observational constraints justify approaches
with more assumed parameters is not clear.)
HCGD argue that the use of a power-law viscosity
can be justified on approximate grounds.
Using the $\alpha$ prescription for the viscosity (Shakura and Sunyaev 1973), 
$\nu = \alpha c_s H \propto {c_s}^2 / \Omega (R) \propto
T(R) R^{3/2}$, where $c_s$ is the sound speed and
$H$ the scale height. If $T(R) \propto R^{-q}$, then
$\nu \propto R^{3/2 - q}$. With $q \sim 1/2$, corresponding
to irradiated disks at large distances from the star (Kenyon and Hartmann 1987;
D'Alessio et al. 1998),
and also found by empirical fitting to apply to most disks in CTTS (Beckwith et al. 1990),
then $\gamma \sim 1$, which is roughly consistent with the
observed slope of the $\mdot$ vs. age data, $\eta \approx 1.5$ (section III.A).

HCGD calculated similarity solutions for viscous evolution for a range of
initial conditions applicable to CCTS disks. Figures 5a
and 5b  show the evolution of mass accretion rate and disk
mass for a subset of models in
this study. Initial disk masses have been taken as 
$M_d(0) = 0.1 \msun$, consistent with small disk
masses remaining after disk-draining episodes of high $\mdot$ (III.C).
Values for other values of $M_d(0)$ can be obtained
by simple scaling.
Model results are shown for three values of the initial
radius, 1, 10, and 100 AU, which cover an order of magnitude
in angular momentum,
consistent with the spread of angular momentum between half
of the binaries in the solar neighborhood (Duquennoy and Mayor 1991).
The calculations assume a temperature
of 10 K at 100 AU, as suggested by  irradiated disk calculations
(D'Alessio et al. 1998), and a central stellar mass of $0.5 \msun$.
The viscosity parameter $\alpha$ 
has been taken as 0.01, except when stated otherwise.
The observed values
of disk masses and mass accretion rates, 
shown in Fig.\ 5a and 5b lie within the region
bounded by the assumed range of initial conditions. 
Disks
with larger initial radii take longer to start evolving,
since the viscous time scale is
$t_{visc} \sim R^2 / \nu \propto R$.

The surface
density of the similarity solutions behaves with
radius as
$$\Sigma \propto { { e^{-(R/R_1)/(t/t_s) }} \over { (R/R_1)} }, \eqno(7)$$
\noindent
where $R_1$ and $t_s$ are characteristic radius and time (see HCGD
for details); it goes like $\propto R^{-1}$ at small radii and
falls sharply at large distances. This last property
determines important
differences in the disk ``sizes'' measured at different wavelengths,
and naturally explains the observed disparity between the optical and mm
sizes. Disk radii measured at millimeter wavelengths are of the order
of a few hundred AU (Dutrey et al. 1996), while the radii
of the disks seen in silhouettes
in the Orion Nebula cluster at 0.6 $\mu$m are much larger, $\sim$ 500 - 1000 AU
(McCaughrean and O'Dell 1996).

Figure 6 shows the predicted radii of models in Fig.\ 5 at
2.7 mm and 0.6$\mu$m, compared to the
observations.
Circles indicate the Dutrey et al. (1996) observations, while the error
bar indicates the range of sizes of the Orion silhouettes.
To calculate the millimeter sizes
the two-dimensional brightness distribution of the disk model
has been convolved with a Gaussian with the appropriate beam size.
Radii at other wavelengths correspond to the radii
where the optical depth is $\sim 1$ at the given wavelength (HCGD).
The theoretical predictions compare well with the observations.
Since dust opacity increases rapidly towards the optical, the
outer tenuous regions of the disk can effectively absorb background
light and produce a large apparent size. In contrast, in the 
millimeter range these outer
cooler, low density regions contribute little to the surface
brightness and the observed sizes are consequently smaller (HCGD).                                              
The larger millimeter size of the  
disk in the binary (open circle) may be indicative of a 
circumbinary disk, for which the present calculations do not apply.

Figure 6 shows the predicted size of one of the
models at 1.87$\mu$m. At an age of
$ \sim 0.5 Myr$, typical of the Orion Nebula cluster,
the infrared sizes are $\sim$ 20\% smaller than
the optical sizes, in good agreement with observations
(McCaughrean et al. 1998). The sharp decline of surface density
with radius predicted by viscously evolving models naturally
explains the observed sizes, without the need for heavily truncated
edges.

Figure 6 also shows with long dashes the predictions for the 
observed sizes in the millimeter range of a disk model
with $\alpha = 0.001$. Since $t_{visc} \sim R^2 / \nu \propto 1 / \alpha$,
disks with small $\alpha$ take much longer to start evolving
and growing, resulting in sizes much smaller than observed
at the typical ages of the young population.
Thus, measurements of disk sizes as a function of time will
place important constraints on the 
characteristic value of $\alpha$ in CTTS disks. 

Disk evolution could be considerably different if angular momentum is 
lost from outer disk regions through a wind (e.g., Pudritz and Norman 
1983, K\"onigl 1989). We have argued elsewhere (Hartmann 1995)
that this is not the case.  Similarly, coagulation of disk material
into bodies that sweep clear the gas will significantly modify this
simple picture of disk evolution.  Nevertheless, it is encouraging
that the observations can be explained with a viscosity ($\alpha \sim 10^{-2}$)
comparable to that estimated in simulations of the Balbus-Hawley
magnetorotational instability (Stone et al. 1996; Brandenburg et al. 1996).

\subsection{E.~~ Effects of companion stars}

There is a large spread in the mass accretion rates and disk masses
as a function of age.  Some of this range is probably
due to uncertainties in age determinations, errors in accretion
rates (for example, due to ignoring inclination effects),
time-variability, and a range in initial conditions. 
However, the potential effects of companion stars cannot be ignored;
since at least $\sim 2/3$ of all systems are binaries (Duquennoy and Mayor 1991),
it is important to consider the effects of a binary companion on the
evolution of disk accretion when comparing with observations.

A companion star may prevent the formation of a disk in its immediate
vicinity, and will try to open gaps on either side of its orbit
(cf. Artymowicz and Lubow 1994, AL; 
Lubow and Artymowicz, this volume).  While the ``initial''
effects of a binary companion can strongly limit disk structure
and accretion, it is important to realize that there may be secondary
effects as well.  Specifically, even with a relatively distant companion,
the inner structure of a viscously-evolving disk will eventually
be affected by the companion, even though the tidal forces are
negligible in this region.  
The reason is that the isolated viscously-evolving
disk can only accrete if its outer regions expand to take up the
necessary angular momentum.  In the similarity solution described
above, the mass accretion rate decreases as a power-law in time,
because as the disk empties out it also expands and thus has
an increasingly long viscous time.  In contrast, if a binary
companion limits the expansion of the disk, once the disk
reaches its maximum size, the viscous time remains constant,
and so the (inner) regions empty out exponentially with time.
(Note that these considerations are relevant only to circumstellar disks,
not circumbinary disks, which can expand.)

Figures 5c and 5d show a very simple calculation of this type of effect,
using the same power-law viscosity used in the standard model,
but now not allowing the disk to expand beyond a certain outer
radius, using the boundary conditions discussed by Pringle (1991).
One observes that when the disk expands to the limiting radius,
the accretion rate first increases slightly, and then drops
precipitously as the disk empties out rapidly.

The significance of this estimate can be seen by noting that
the median binary separation is roughly 30 AU (Duquennoy and Mayor 1991).
With the reference disk model used, and estimating
a truncation radius $\sim 1/3$ of the binary separation (AL),
this would mean that even if all binaries originally had circumstellar
disks, half of those binaries would have their disks empty out
by an age of 1 Myr.  These estimates are roughly consistent with
the percentage of $\sim$ 1 Myr WTTS
in Taurus, $\sim$ 45 \% (Kenyon and Hartmann 1995); the predicted
fraction of WTTS could be even higher if 
indeed the fraction of binaries in Taurus is higher than
in the solar neighborhood (Simon et al. 1995).

These estimates of binary effects on disk evolution
are rough, and the model
ignores the effects of the (significant) eccentricities of binary
orbits (AL), but they serve to illustrate the importance
of identifying stellar companions to understand disk evolution
in individual systems.  
Figures 5c and d do not show a very strong correlation of mass accretion
rates with binarity, though there is a significant effect on disk
masses (cf. Jensen et al. 1994; Mathieu et al. 1995; Osterloh and Beckwith 1995).  
In any case, many of the systems
shown have not been studied carefully for potential companions, and
in general much work remains to be done in this area.

\mainsection{IV. ~~SUMMARY AND IMPLICATIONS}

Figure 7 summarizes the ideas presented so far in a sketch of disk
evolution with time for a single star. After maybe a short initial period of high
mass accretion rate, the disk remains most of the time 
in a quiescent state.
Episodes of high mass accretion rate are triggered mostly during the
phase where the disk is still immersed in the infalling envelope,
in which we expect most of the star to be built.
After the infall ceases and the star emerges as
a T Tauri star, the disk evolves viscously,
$\mdot$ slowly decreases with time,
at the same time that the disk expands and its mass decreases.

There are several implications of these results for star and planet formation.
First, disk accretion during the protostar phase appears to be
highly variable.  This may call into question theories of the birthline,
or the initial position of stars in the HR diagram (Stahler 1988;
Hartmann et al. 1997), which
assume steady accretion at the rates of infall of the protostellar envelope.
It is conceivable that planetesimals or other bodies form in the disk during
this phase but are swept into the star as the disk accretes, perhaps partly
accounting for some of the accretion variability - our observations
really only probe energy release in the inner disk, near the star.
Second, there appears to be a wide variety of disk masses and accretion rates
(say, a range of an order of magnitude) during the T Tauri phase, 
produced in part by differences in initial angular momenta.  This
may mean that any consequent planetary systems which form could have
quite different properties.  Third, the ability of viscous accretion disk
models to explain the observations so far suggests that substantial migration
of material occurs during the T Tauri phase; this migration as disks actively
accrete may be important in explaining some of the extrasolar planets which
lie close to their star.  Fourth, the presence of binary companions obviously
does not always prevent disk formation, but they may accelerate 
(circumstellar) disk accretion.  Improved mm and submm interferometry, as well
as infrared speckle searches for companion stars and improved radial velocity
studies to search for close, low-mass stellar companions, will lead to a
greatly improved understanding of disk evolution.

\subsection{Acknowledgements}
We thank a number of people for useful discussions and valuable
insight, including James Muzerolle, Erik Gullbring, Paola D'Alessio,
Cesar Brice\~no, Ray Jayawardhana, Suzan Edwards, Lynne Hillenbrand, Michael Meyer,
Bo Reipurth, and David Wilner.
This work was supported in part by NASA grant NAG5-4282.


\vskip .5in
\centerline{\bf REFERENCES}
\vskip .25in

\ref{Adams, F. C. and Shu, F. H. 1986. Infrared Spectra
of Rotating Protostars. {\refit Astrophys.\ J.\/}  308:836-853.}

\ref{Andr\'e, P., Ward-Thompson, D., and Barsony, M. 1993.
Submillimeter continuum observations of Rho Ophiuchi A - The candidate
protostar VLA 1623 and prestellar clumps.
{\refit Astrophys.\ J.\/}  406:122-141.}

\ref{Artymowicz, P. and Lubow, S. H. 1994,
Dynamics of binary-disk interaction. 1: Resonances and disk gap size.
{\refit Astrophys.\ J.\/}  421:651-667.}

\ref{Basri, G., Marcy, G.W. and Valenti, J. A. 1992.
Limits on the magnetic flux of pre-main sequence stars.
{\refit Astrophys.\ J.\/}  390:622-633

\ref{Beckwith, S. V. W., Sargent, A. I., Chini, R. S.
and Guesten, R. 1990.
A survey for circumstellar disks around young stellar objects.
{\refit Astron.\ J.\/}  99:924-945.}

\ref{Bell, K. R. and Lin, D. N. C. 1994.
Using FU Orionis outbursts to constrain self-regulated protostellar disk models.
{\refit Astrophys.\ J.\/}  427:987-1004.}

\ref{Bell, K. R., Lin, D. N. C, Hartmann, L. and Kenyon, S. J. 1995.
The FU Orionis outburst as a thermal accretion event: Observational
constraints for protostellar disk models.
{\refit Astrophys.\ J.\/}  444:376-395.}

\ref{Bertout, C. 1989.
T Tauri stars - Wild as dust.
{\refit Ann.\ Rev.\ Astron.\ Astrophys.\/}  27:351-395.}

\ref{Bonnell, I. and Bastien, P. 1992,
A binary origin for FU Orionis stars.
{\refit Astrophys.\ J.\/}  401:L31-L34.}

\ref{ Brandenburg, A., Nordlund, A., Stein, R.F., and Torkelsson, U. 1996. 
{\refit Astrophys.\ J.\/}  458:L45.}

\ref{Calvet, N., Pati\~no, A., Magris C., G. and D'Alessio, P. 1991.
Irradiation of accretion disks around young objects. I - Near-infrared CO
bands. {\refit Astrophys.\ J.\/}  380:617-630.}

\ref{Calvet, Magris C., G., Pati\~no, A. and D'Alessio, P. 1992.
Irradiation of Accretion Disks Around Young Objects. II.
Continuum Energy Distribution.
{\refit Rev. Mex. Astron. Astrofis.\/}  24:27-42.}

\ref{Calvet, N. and Hartmann, L. 1992.
Balmer line profiles for infalling T Tauri envelopes.
{\refit Astrophys.\ J.\/}  386:239-247.}

\ref{Calvet, N., Hartmann, L. and Strom, S. E. 1997.
Near-Infrared Emission of Protostars.
{\refit Astrophys.\ J.\ Lett.\/}  481:912-917.}

\ref{Calvet, N. 1995
Properties of the Winds of T Tauri Stars. In
{\refit Herbig-Haro Flows and the Birth of Stars; IAU Symposium No. 182 \/}, eds. 
B. Reipurth and C. Bertout. (Dordrecht: Kluwer Academic Publishers), p.
417-432.}

\ref{Calvet. N. and Gullbring, E. 1998, {\refit Astrophys.\ J.\/}  (in press)

\ref{Chiang, E. I. and Goldreich, P. 1997.
Spectral Energy Distributions of T Tauri Stars with Passive Circumstellar Disks
{\refit Astrophys.\ J.\/}  490:368-376.}

\ref{Clarke, C.J., Lin, C
D.N.C. and Pringle, J. E. 1990.
Pre-conditions for disc-generated FU Orionis outbursts.
{\refit Mon.\ Not.\ Roy.\ Astron.\ Soc.\/}  242:439-446.}

\ref{D'Alessio, P., Cant\'o, J., Calvet, N. and Lizano, S. 1998.
Accretion Disks around Young Objects. I. The Detailed Vertical Structure.
{\refit Astrophys.\ J.\/}  500:411-427.}

\ref{D'Antona, F. and Mazitelli, I. 1994.
New pre-main-sequence tracks for M less than or equal to 2.5 solar mass as
tests of opacities and convection model.
{\refit Astrophys.\ J. Suppl.\/} 90:467-500.}

\ref{Duquennoy, A. and Mayor, M. 1991.
Multiplicity among solar-type stars in the solar neighbourhood. II -
Distribution of the orbital elements in an unbiased sample.
{\refit Astron.\ Astrophys.\/}  248:485-524.}
 
\ref{Dutrey, A., Guilloteau, S., Duver, G., Prato, L., Simon, M.,
Schuster, K. and Menard, F. 1996.
Dust and gas distribution around T Tauri stars in Taurus-Auriga. I.
Interferometric 2.7mm continuum and $CO_{13}$ J=1-0 observations.
{\refit Astron.\ Astrophys.\/}  309:493-504.} 
 
\ref{Edwards, S., Hartigan, P., Ghandour, L. and Andrulis, C. 
Spectroscopic evidence for magnetospheric accretion in classical T Tauri stars.
{\refit Astron.\ J.\/}  108:1056-1070.}

\ref{Elias, J. H. 1978. A study of the IC 5146 dark cloud complex.
{\refit Astrophys.\ J.\/}  223:859-875.}

\ref{Gauvin, L. S. and Strom, K. M. 1992.
A study of the stellar population in the Chamaeleon dark clouds.
{\refit Astrophys.\ J.\/}  385:217-231.}

\ref{\ref{Graham, J.A. and Frogel, J. A. 1985. 
An FU Orionis star associated with Herbig-Haro object 57.
{\refit Astrophys.\ J.\/}  289:331-341.}

\ref{Greene, T. P. and Lada, C. J. 1996.
Near-Infrared Spectra and the Evolutionary Status of Young Stellar Objects:
Results of a 1.1-2.4 micron Survey.
{\refit Astron.\ J.\/} 112:2184-2221.}

\ref{Greene, T. P. and Meyer, M. R. 1995.
An Infrared Spectroscopic Survey of the rho Ophiuchi Young Stellar Cluster:
Masses and Ages from the H-R Diagram.
{\refit Astrophys.\ J.\/}  450:233-244.}

\ref{Gullbring, E., Hartmann, L., Brice\~no, C. and Calvet, N. 1998
Disk Accretion Rates for T Tauri Stars.
{\refit Astrophys.\ J.\/}  492:323-341.}

\ref{Hanson, M. M. and Conti, P. S. 1995.
Identification of Ionizing Sources and Young Stellar Objects in M17.
{\refit Astrophys.\ J.\ Lett.\/}
448:L45-L48.}

\ref{Hartigan, P., Strom, S. E., Edwards, S., Kenyon, S. J., Hartmann, L.,
Stauffer, J. and Welty, A. D. 1991.
Optical excess emission in T Tauri stars.
{\refit Astrophys.\ J.\/}  382:617-635.}

\ref{Hartigan, P., Edwards, S. and Ghandour, L. 1995.
Disk Accretion and Mass Loss from Young Stars.
{\refit Astrophys.\ J.\/}  452:736-768.}

\ref{Hartmann, L., Hewett, R. and Calvet, N. 1994.
Magnetospheric accretion models for T Tauri stars. 1: Balmer line profiles
without rotation.
{\refit Astrophys.\ J.\/}  426:669-687.}

\ref{Hartmann, L. 1995.
Observational Constraints on Disk Winds. In
{\refit Circumstellar Disks, Outflows and Star Formation}.  
{\refit Rev. Mexicana Astron. Astrof. Serie de Conferencias.\/}
1:285-291.}

\ref{Hartmann, L. and Kenyon, S. J. 1996. The FU Orionis phenomenon.
{\refit Annu. Rev. Astron. Astrophys.\/}  34:207-240.}

\ref{Hartmann, L., Cassen, P. and Kenyon, S. J. 1997.
Disk Accretion and the Stellar Birthline. 
{\refit Astrophys.\ J.\/}  475:770-785.}

\ref{Hartmann, L., Calvet, N., Gullbring, E. and D'Alessio, P. 1988.
Accretion and the Evolution of T Tauri Disks.
{\refit Astrophys.\ J.\/}  495:385-400.}

\ref{Herbig, G. H. 1977. Eruptive phenomena in early stellar evolution.
{\refit Astrophys.\ J.\/}  217:693-715.} 

\ref{Hodapp, K-W, Hora, J. L., Rayner, J. T., Pickles, A. J., and Ladd, E. F., 1996.
An outburst of a deeply embedded star in Serpens.
{\refit Astrophys.\ J.\/}  468:861-870.}

\ref{Jensen, E. L. N., Mathieu, R. D. and Fuller, G. A. 1994. 
A connection between submillimeter continuum flux and separation in young
binaries.
{\refit Astrophys.\ J.\/}  429:L29-L32.}

\ref{Kawazoe, E. and Mineshige, S. 1993.
Unstable accretion disks in FU Orionis stars.
{\refit Publ.\ Astron.\ Soc.\ Pacific}  45:715-725.}

\ref{Kenyon, S. J. and Hartmann, L. 1987.
Spectral energy distributions of T Tauri stars - Disk flaring and limits on
accretion.
{\refit Astrophys.\ J.\/}  323:714-733.}

\ref{Kenyon, S. J. and Hartmann, L. 1995.
Pre-Main-Sequence Evolution in the Taurus-Auriga Molecular Cloud.
{\refit Astrophys.\ J. Suppl.\/} 101:117-171.}

\ref{Kenyon, S. J. and Hartmann, L. 1996.
The FU Orionis Phenomenon.
{\refit Ann.\ Rev.\ Astron.\ Astrophys.\/}  34:207-240.}

\ref{Kenyon, S. J., Hartmann, L., Strom, K. M. and Strom, S. E. 1990.
An IRAS survey of the Taurus-Auriga molecular cloud.
{\refit Astron.\ J.\/} 99:869-887.}

\ref{Kenyon, S.J., Calvet, N., and Hartmann, L.  1993a.
The embedded young stars in the Taurus-Auriga molecular cloud. I.
Models for the spectral energy distribution.
{\refit Astrophys.\ J.\/}  414:676-694.}

\ref{Kenyon, S.J., Whitney, B., A., Gomez, M., and Hartmann, L.  1993b.
The embedded young stars in the Taurus-Auriga molecular cloud. II - Models
for scattered light images.
{\refit Astrophys.\ J.\/}  414:773-792.}

\ref{Kenyon, S., Gomez, M., Marzke, R. O. and Hartmann, L. 1994.
New pre-main-sequence stars in the Taurus-Auriga molecular cloud.
{\refit Astron.\ J.\/} 108:251-261.}

\ref{Kenyon, S., J., Yi, I., and Hartmann, L. 1996.
A Magnetic Accretion Disk Model for the Infrared Excesses of T Tauri Stars.
{\refit Astrophys.\ J.\/}  462:439-455.}

\ref{K\"onigl, A. 1989.
Self-similar models of magnetized accretion disks.
{\refit Astrophys.\ J.\/}  342:208-223.}
 
\ref{K\"onigl, A. 1991.
Disk accretion onto magnetic T Tauri stars.
{\refit Astrophys.\ J.\ Lett.\/}  370:L39-L43.}
 
\ref{Larson, R. B. 1984.
Gravitational torques and star formation.
{\refit Mon.\ Not.\ Roy.\ Astron.\ Soc.\/}  206:97-207.}

\ref{Lin, D. N. C. and Bodenheimer, P. 1982.
On the evolution of convective accretion disk models of the primordial solar
nebula.
{\refit Astrophys.\ J.\/}  262:768-779.}

\ref{Lin, D. N. C. and Papaloizou, J. C. B. 1985.
On the dynamical origin of the solar system.
In {\refit Protostars \& Planets {I}{I}\/}, eds.\
D. C. Black and M. S. Matthews
(Tucson: Univ.\ of Arizona Press), pp.\ 981-107.}

\ref{Lynden-Bell, D. and Pringle, J. E. 1974
The evolution of viscous discs and the origin of the nebular variables.
{\refit Mon.\ Not.\ Roy.\ Astron.\ Soc.\/} 168:603-638.}

\ref{Mathieu, R., Adams, F. C., Fuller, G. A., Jensen, E. L., N., Koerner, D. W.
and Sargent, A. 
Submillimeter Continuum Observations of the T Tauri Spectroscopic Binary GW Orionis.
{\refit Astron.\ J.\/}  109:2655-2669.}

\ref{McCaughrean, M. J. and O'Dell, C. R. 1996
Direct Imaging of Circumstellar Disks in the Orion Nebula.
{\refit Astron.\ J.\/}  111:1977-1986.}
 
\ref{McCaughrean, M. J., Chen, H., Bally, J., Erickson, Ed, Thompson, R.,
Rieke, M., Schneider, G., Stolovy, S. and Young, E. 1998.
High-resolution near-infrared imaging of the Orion 114-426
silhouette disk.
{\refit Astrophys.\ J.\ Lett.\/} 492:L157-161.}
 
\ref{Meyer, M. R., Calvet, N. and Hillenbrand, L. A. 1997.
Intrinsic Near-Infrared Excesses of T Tauri Stars: Understanding the Classical T
Tauri Star Locus.
{\refit Astrophys.\ J.\/}  114:288-300.}

\ref{Muzerolle, J., Calvet, N. and Hartmann, L. 1998a.
Magnetospheric Accretion Models for the Hydrogen Emission Lines of T Tauri
Stars. {\refit Astrophys.\ J.\/}  492:743-753.}

\ref{Muzerolle, J., Hartmann, N. and Calvet, N. 1998b.
Emission-Line Diagnostics of T Tauri Magnetospheric Accretion. I. Line Profile
Observations. {\refit Astron.\ J.\/}  116:455-468.}

\ref{Muzerolle, J., Hartmann, N. and Calvet, N. 1998c.
A Br$\gamma$ Probe of Disk Accretion in T Tauri Stars
and Embedded Young Stellar Objects.
{\refit Astrophys.\ J.\/}  (in press).}

\ref{Moriarty-Schieven, G.H. and Snell, R. 1988. High-resolution images
of the L1551 molecular outflow. II. Structure and kinematics. {\refit Astrophys.\ J.\/}  332:364-378.}

\ref{Osterloh, M. and Beckwith, S. V. W. 1995.
Millimeter-wave continuum measurements of young stars.
{\refit Astrophys.\ J.\/}  439:288-302.}

\ref{Pringle, J. E. 1981.
Accretion discs in astrophysics.
{\refit Ann.\ Rev.\ Astron.\ Astrophys.\/}  19:137-162.}

\ref{Pringle, J. E. 1991.
The properties of external accretion discs.
{\refit Mon.\ Not.\ Roy.\ Astron.\ Soc.\/}  248: 754-759.}

\ref{Pudritz, R. E. and Norman, C. A. 1983.
Centrifugally driven winds from contracting molecular disks.
{\refit Astrophys.\ J.\/} 274:677-697.}

\ref{Reipurth, B. 1985.
Herbig-Haro objects and FU Orionis eruptions - The case of HH 57.
{\refit Astron.\ Astrophys.\/}  143:435-442.}

\ref{Reipurth, B. and Aspin, C. 1997. Infrared spectroscopy of Herbig-Haro
energy sources. {\refit Astron.\ J.\/}} 114:2700-2707.}

\ref{Rucinski, S. M. 1985,
Iras observations of T Tauri and post-T Tauri stars.
{\refit Astron.\ J.\/}  90:2321-2330.}

\ref{Rydgren, A. E. and Zak, D. S. 1987.
On the spectral form of the infrared excess component in T Tauri systems.
{\refit Publ.\ Astron.\ Soc.\ Pacific} 99:141-145.}

\ref{Sandell, G. and Aspin, C. 1998. PP13S, a young, low-mass FU
Orionis-type pre-main sequence star. {\refit Astron. Astrophys.\/}  333:1016-1025.}

\ref{Shakura, N. I. and Sunyaev, R. A. 1973.
Black holes in binary systems. Observational appearance.
{\refit Astron.\ Astrophys.\/}  24:337-355.}

\ref{Shu, F. H., Adams, F. C. and Lizano, S. 1987.
Star formation in molecular clouds - Observation and theory.
{\refit Ann.\ Rev.\ Astron.\ Astrophys.\/}  25:23-81.}

\ref{Shu, F. H., Najita, J., Ostriker, E., Wilkin. F., Ruden, S. P. and Lizano, S. 1994.
Magnetocentrifugally driven flows from young stars and disks. 1: A
generalized model.
{\refit Astrophys.\ J.\/}  429:781-796.}

\ref{Simon, M., Ghez, A. M., Leinert, C. H., Cassar, L., Chen, W. P.,
Howell, R. R., Jameson, R. F., Matthews, K., Neugebauer, G. and Richichi, A. 1995.
A lunar occultation and direct imaging survey of multiplicity in the Ophiuchus
and Taurus star-forming regions.
{\refit Astrophys.\ J.\/}  443:625-637.}

\ref{Stahler, S. W. 1988.
Deuterium and the stellar birthline.
{\refit Astrophys.\ J.\/}  332:804-825.}

\ref{Staude, H. J. and Neckel, T. 1991.
RNO 1B - A new FUor in Cassiopeia.
{\refit Astron.\ Astrophys.\/}  244: L13-L16.}

\ref{Stepinski, T. F. 1998.
Diagnosing Properties of Protoplanetary Disks from their evolution.
{\refit Astrophys.\ J.\/}  in press.}

\ref{Stone, J.M., Hawley, J.F., Gammie, C. F., and Balbus, S.A. 1996. 
{\refit Astrophys.\ J.\/}  463:656.}

\ref{Valenti, J. A., Basri, G. and Johns, C. M. 1993
Tauri stars in blue.
{\refit Astron.\ J.\/}  106:2024-2050.}

%
%


\vfill\eject

\vskip1cm
\vbox{
\centerline{
\hbox{
 \epsfxsize 15truecm\epsffile{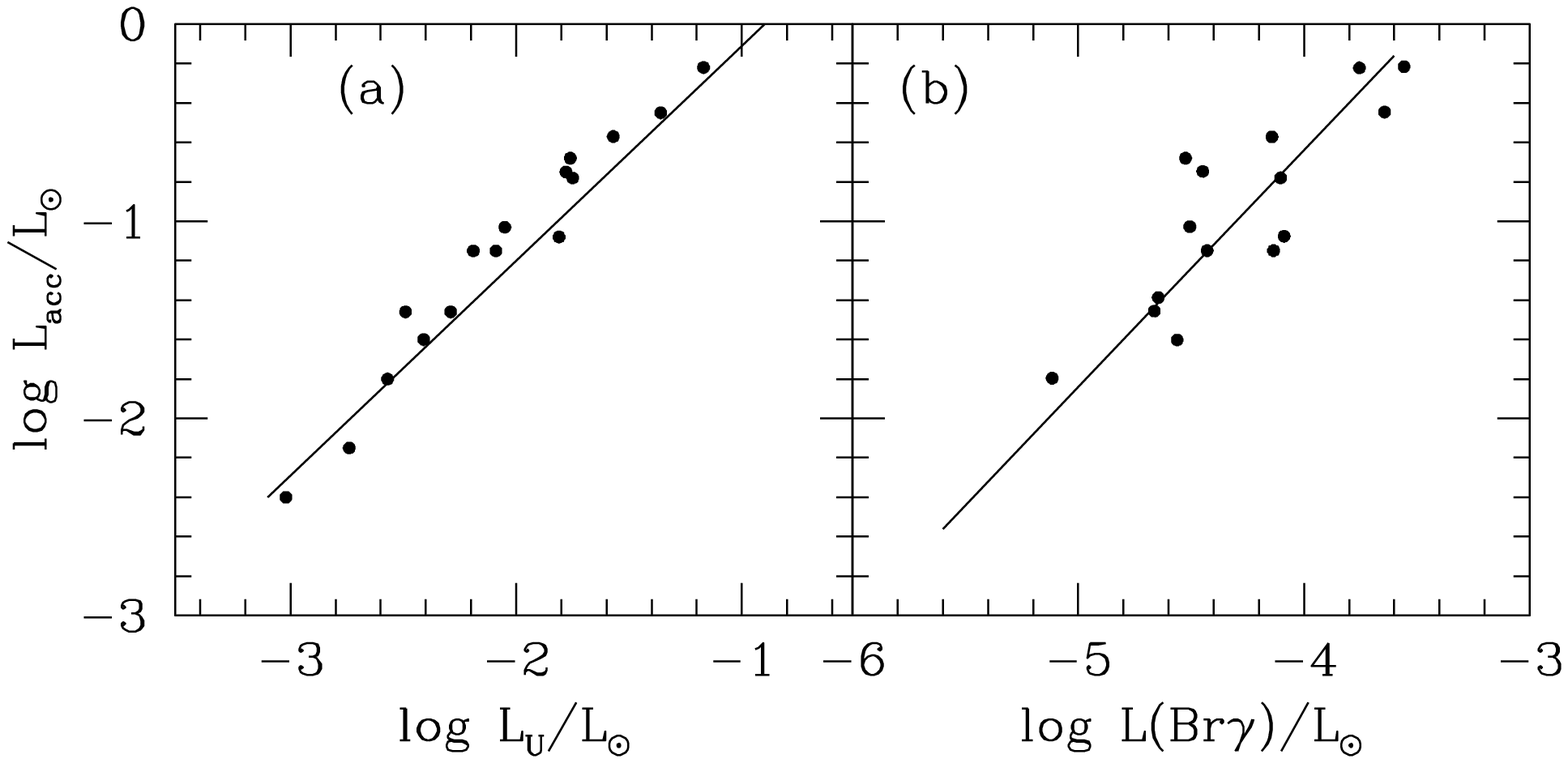}
}}
\vskip 0.5 truecm
\noindent {Figure 1.
\capskip Relationship between accretion
luminosity and (a) the excess luminosity in the U band, and (b) the
luminosity in Br$\gamma$ for a sample of CTTS in Taurus. Data from
Gullbring et al. (1997) and Muzerolle et al. (1998c).}
}
\vglue 0.2 truecm
}

\vfill\eject
 
\vskip1cm
\vbox{
\centerline{
\hbox{
 \epsfxsize 15truecm\epsffile{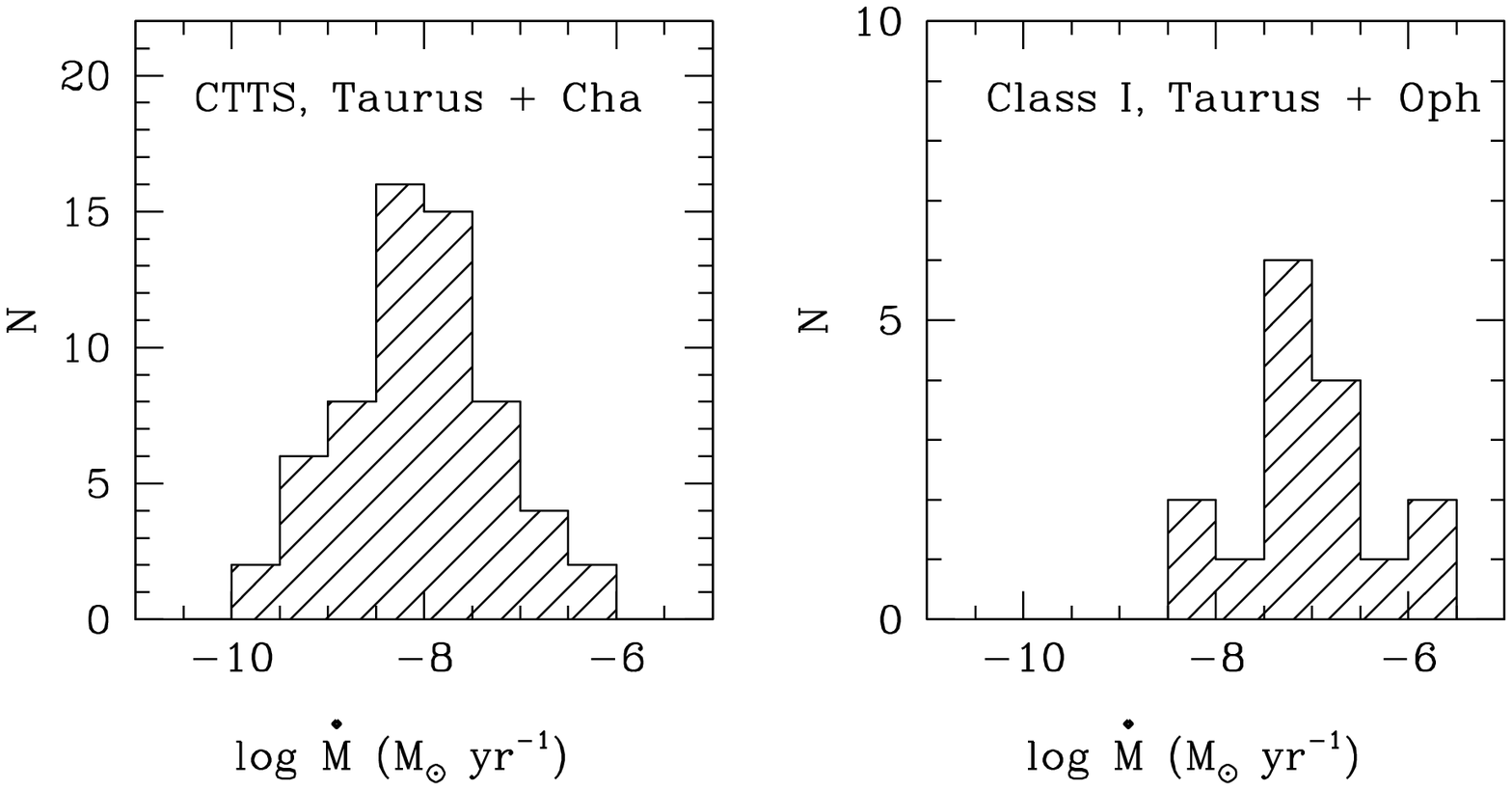}
}}
\vskip 0.5 truecm
\noindent {Figure 2.
\capskip Histogram showing the distribution
of mass accretion rates. (a) For T Tauri stars in Taurus and Cha I,
with $\dot M$ determined from blue spectra or U
magnitudes, and
(b) For Class I sources in Taurus and $\rho$ Oph,
determined from Br$\gamma$ measurements.}
\vglue 0.2 truecm
}

\vfill\eject
\vskip1cm 
\vbox{  
\centerline{ 
\hbox{ 
 \epsfxsize 15truecm\epsffile{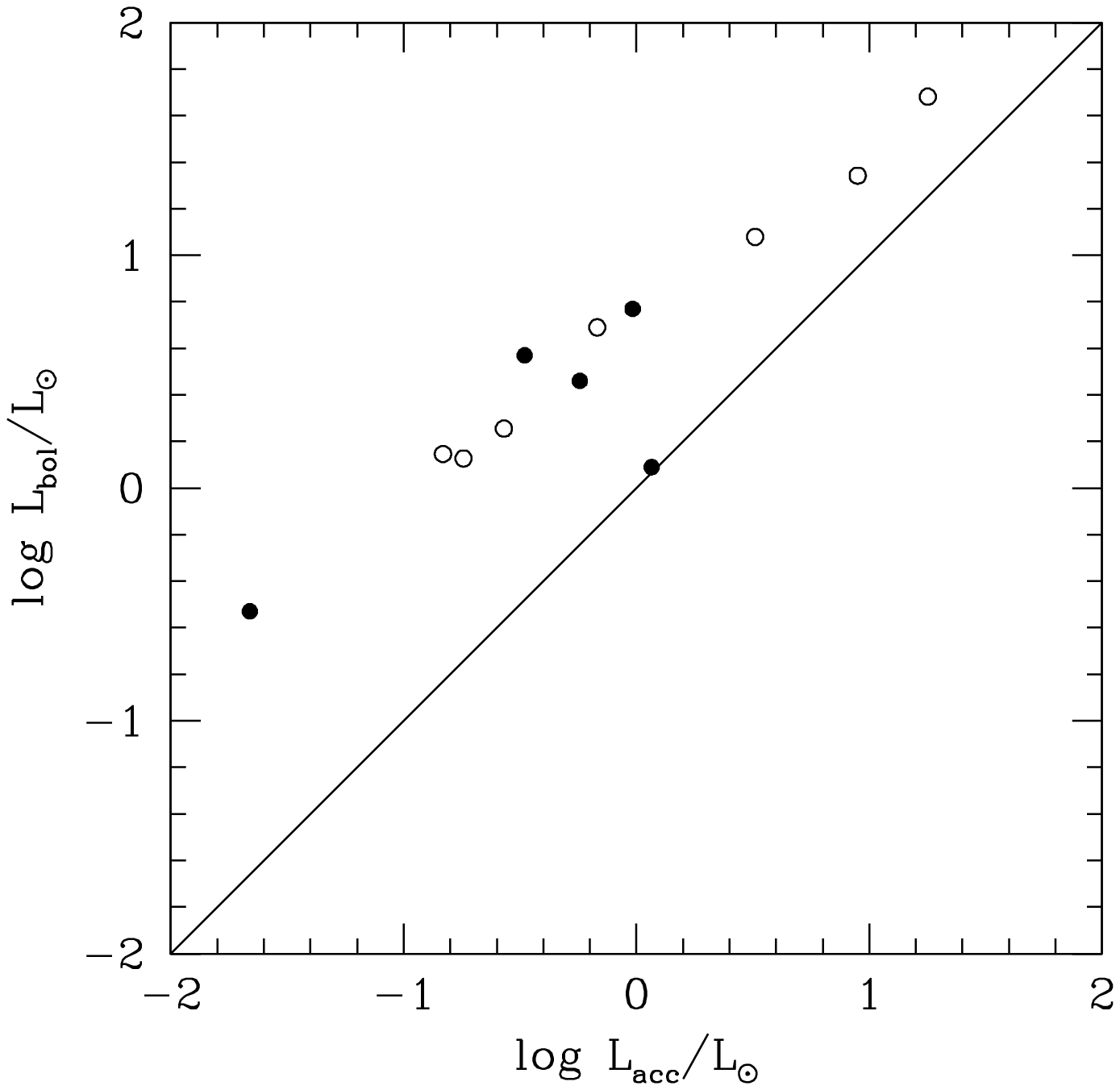} 
}}
\vskip 0.5 truecm 
\noindent {Figure 3. 
\capskip Relationship between accretion luminosity and
bolometric luminosity for Class I sources in Taurus (filled circles) and in
$\rho$ Oph (open circles)
(from Muzerolle et al. 1998c).}
\vglue 0.2 truecm 
} 

\vfill\eject
\vskip1cm 
\vbox{   
\centerline{  
\hbox{  
 \epsfxsize 15truecm\epsffile{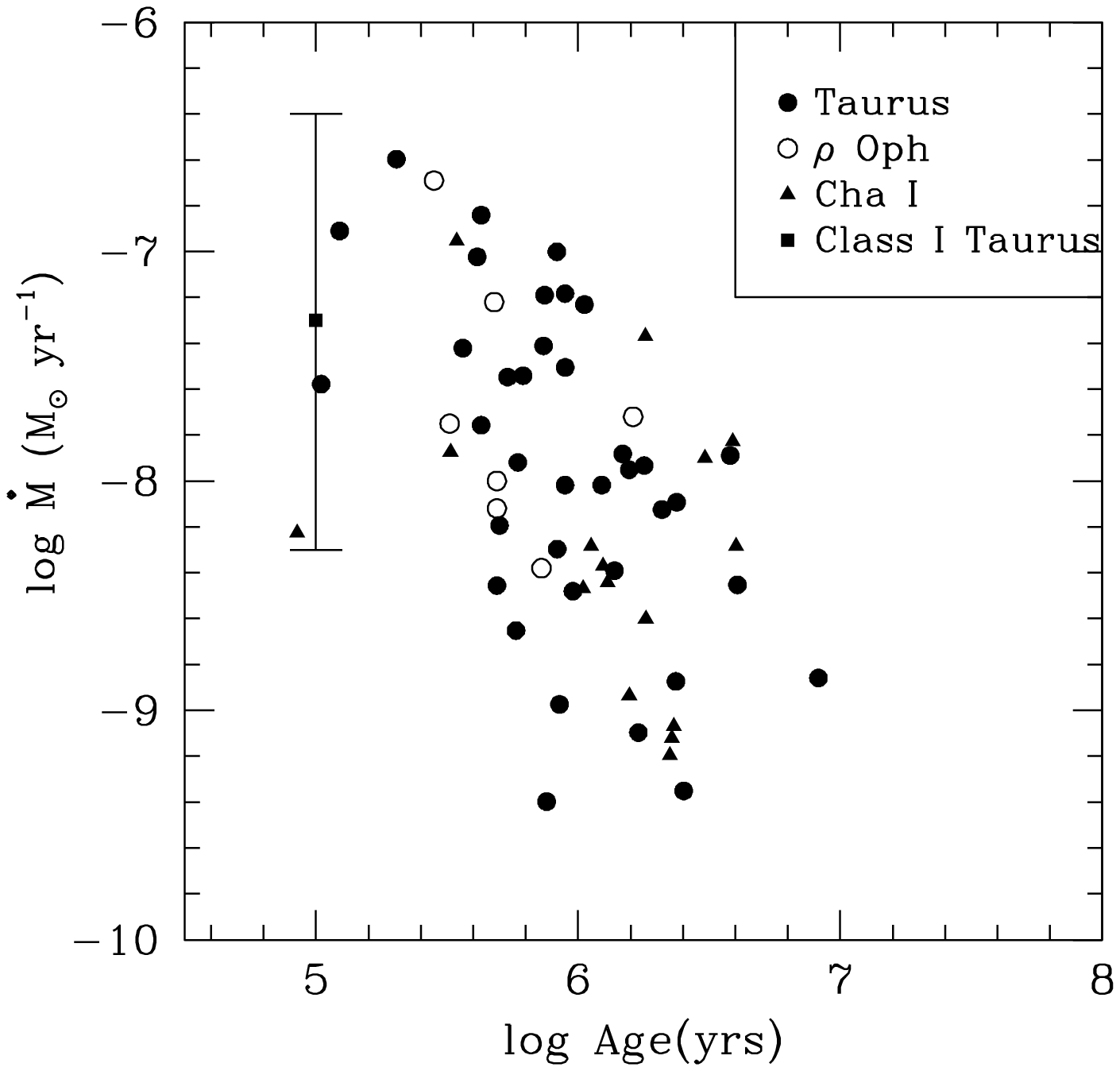}  
}} 
\vskip 0.5 truecm  
\noindent {Figure 4.  
Observed mass accretion rate vs. age for
CTTS in Taurus, Cha I, and $\rho$ Oph.
Mass accretion rates have been
obtained by the methods described in section II.
Ages for the CTTS have been estimated
from the position in the HR diagram and comparison with
the theoretical tracks from D'Antona and Mazitelli (1994,
CMA case). Luminosities and spectral types were taken
from Kenyon and Hartmann (1995) for Taurus, Gauvin and Strom (1992)
for Cha I, and Greene and Meyer (1995) for $\rho$ Oph.
The mean and dispersion of the
estimated mass accretion rates for Class I sources is also shown for
comparison (see II.C).
The mean age for the Class I sources is assumed to be
0.1 Myr.}
\vglue 0.2 truecm  
}

\vfill\eject
\vskip1cm 
\vbox{   
\centerline{  
\hbox{  
 \epsfxsize 15truecm\epsffile{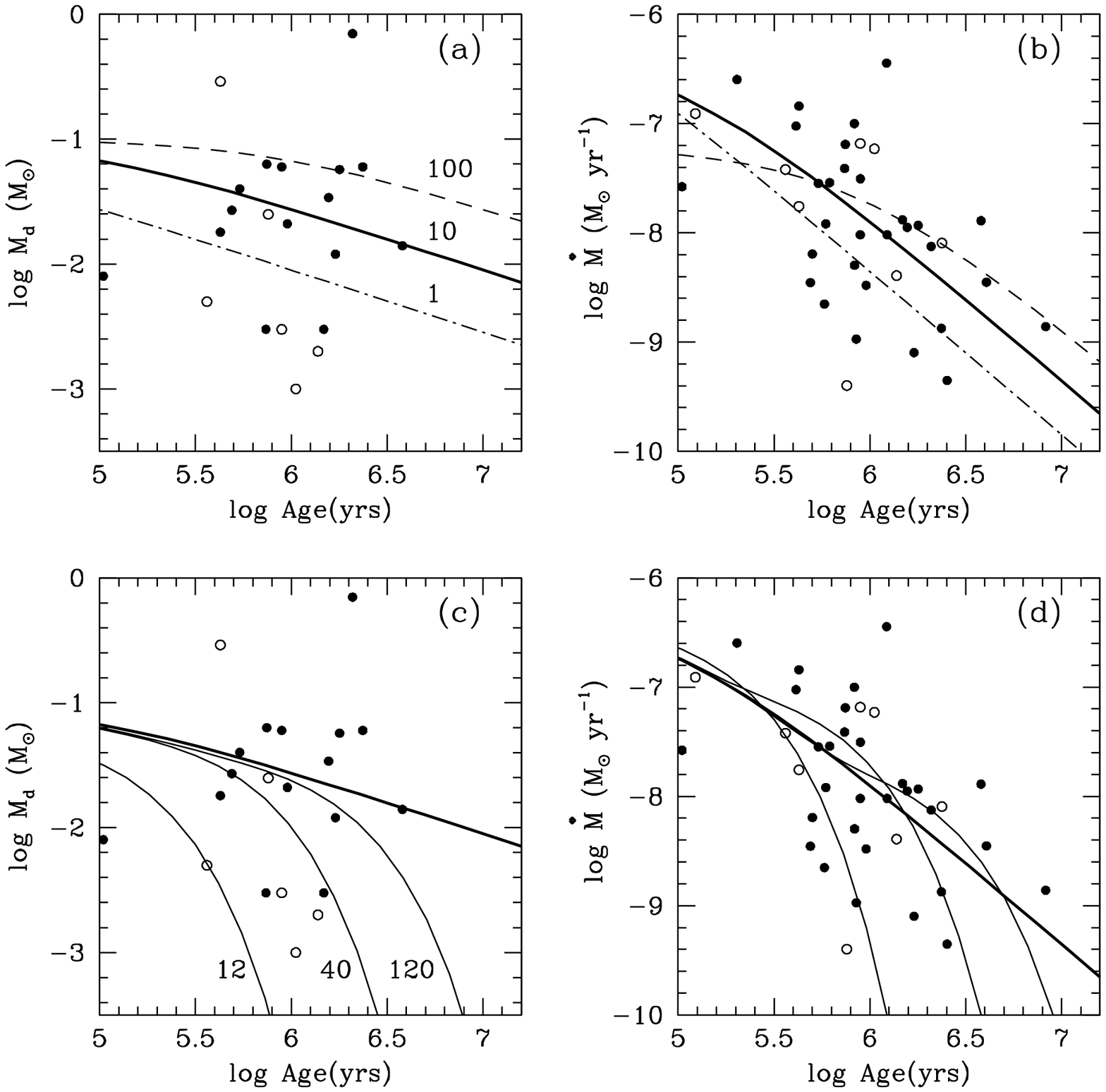}  
}} 
\vskip 0.5 truecm  
\noindent {Figure 5.  
\capskip Similarity solution for disk evolution,
with $\nu \propto R$, compared to observations. Upper panels:
Evolution for isolated disks. (a)  Disk mass vs. time.  (b) $\mdot$ vs. time.
Models shown have initial disk
mass of 0.1 $\msun$ and initial disk radii (marked):
1, 10, and 100 AU. Lower panels: Evolution
with binary companions. (c)  Disk mass vs. time. (d) $\mdot$ vs. time.
Models are shown for initial disk
mass and radius of 0.1 $\msun$ and 10 AU,
and for three binary separations: 30 AU, 100 AU,
and 300 AU (corresponding to truncation radii of 12, 40, and 120 AU, marked).
The corresponding evolution for the isolated disk is shown for
comparison (heavy line).
Binaries indicated by open circles.}
\vglue 0.2 truecm  
}

\vfill\eject 
\vskip1cm  
\vbox{   
\centerline{  
\hbox{  
 \epsfxsize 15truecm\epsffile{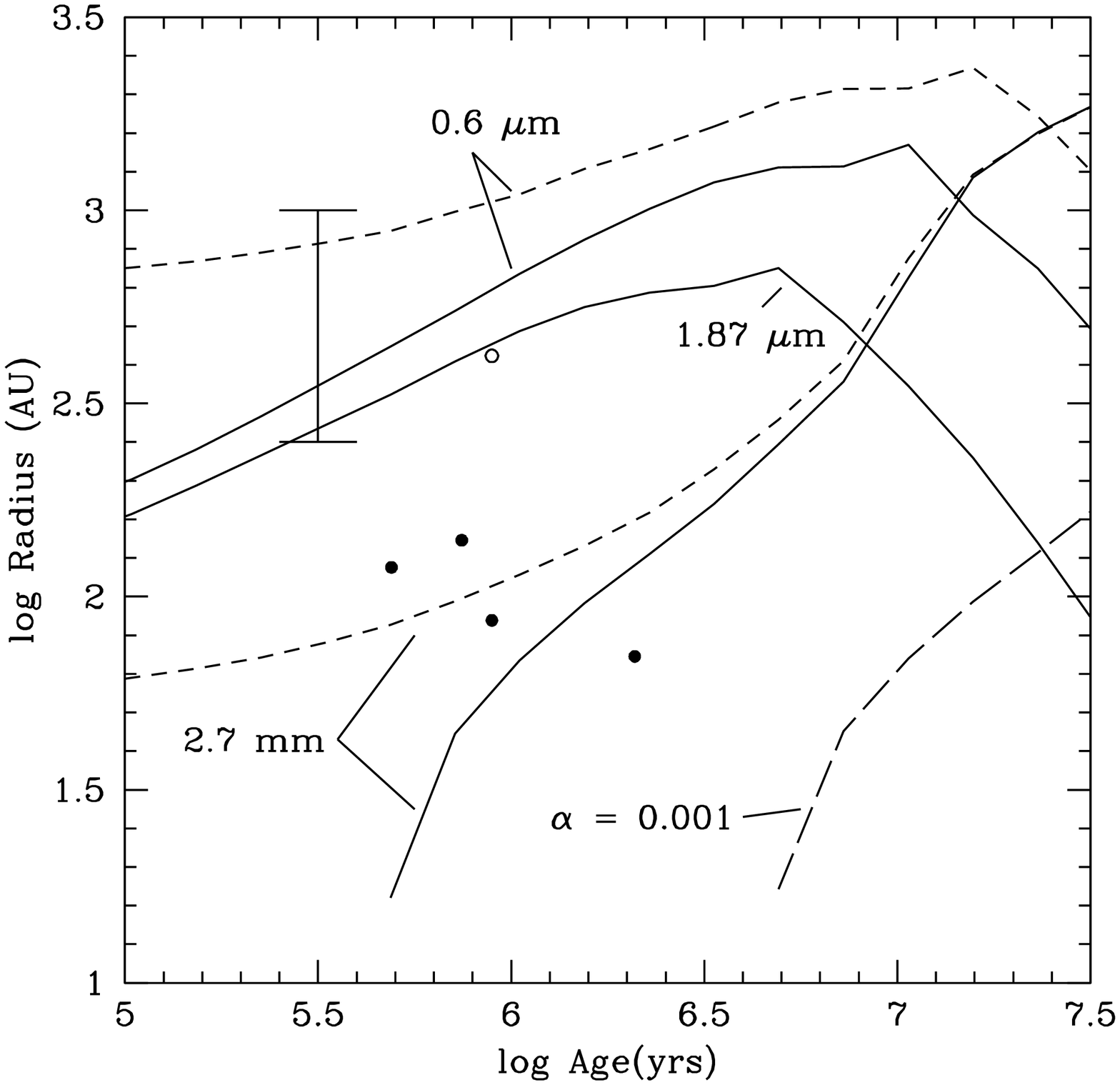}   
}}  
\vskip 0.5 truecm   
\noindent {Figure 6.
\capskip Characteristic disk sizes for viscous
evolution  as observed at 0.6 $\mu$m, 1.87 $\mu$m, and 2.7 mm. Models
are shown for an initial mass of 0.1 $\msun$ and
initial radii 10 AU (solid) and 100 AU (short dashes).
A model with initial radii 10 AU but $\alpha = 0.001$
is shown for comparison (long dashes).
Data from Dutrey et al. (1996) are shown as circles,
and the error bar indicates the range of
sizes measured in the disks seen in silhouettes
in the Orion Nebula cluster (McCaughrean and O'Dell 1996).
Binaries are indicated by open circles. See text.}
\vglue 0.2 truecm
}

\vfill\eject  
\vskip1cm  
\vbox{   
\centerline{   
\hbox{   
\epsfxsize 15truecm\epsffile{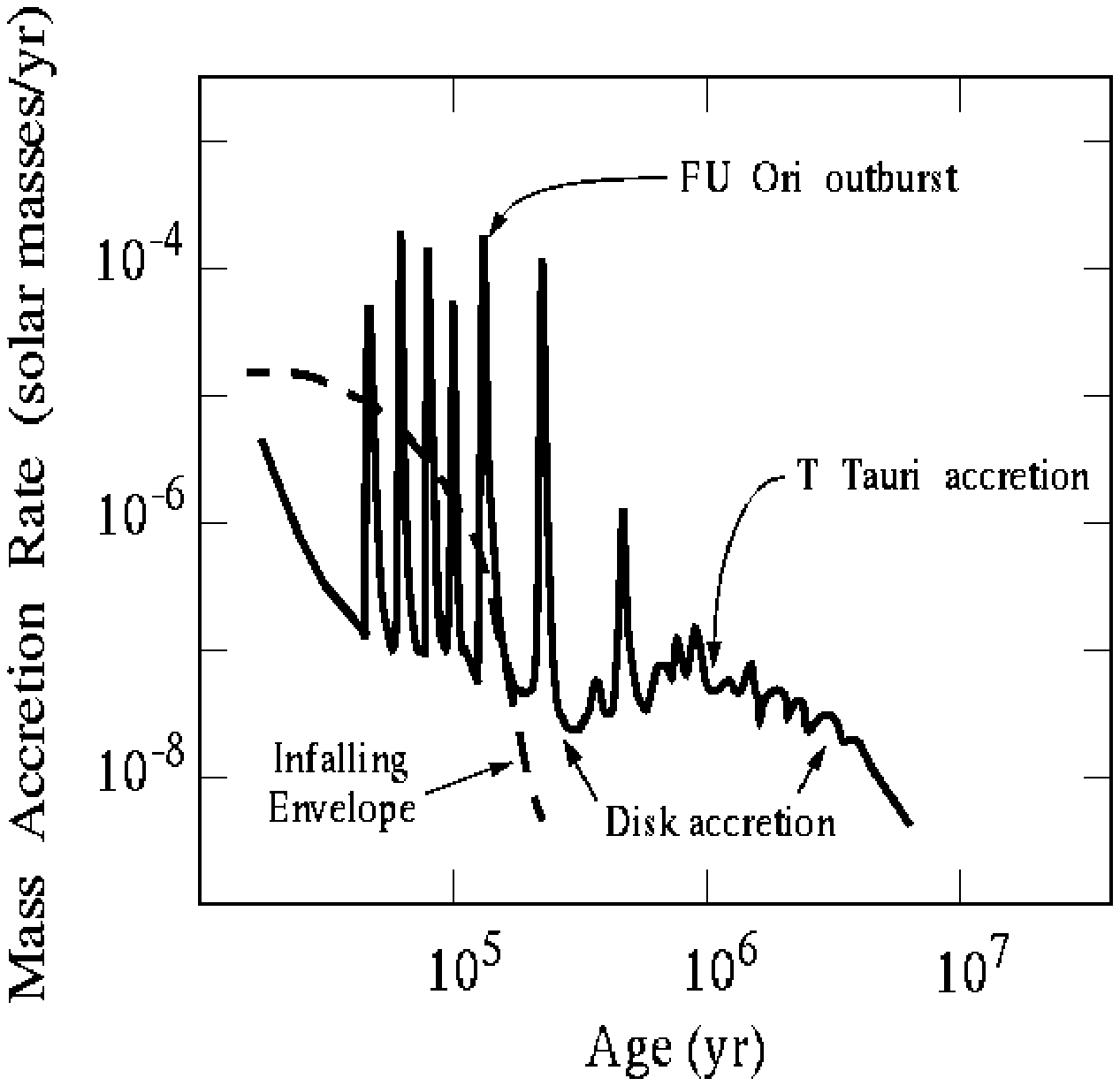}
}}   
\vskip 0.5 truecm    
\noindent {Figure 7.
\capskip Sketch of disk evolution with time,
summarizing the ideas presented in this chapter. The disk remains most of
the time in a quiescent state, punctuated by episodes of
high $\mdot$ as long as the envelope feeds mass to the disk.
When infall ceases, the disk evolves viscously
and $\mdot$ slowly decreases with time.}
\vglue 0.2 truecm 
}

\bye